\documentclass[conference]{IEEEtran}

\IEEEoverridecommandlockouts

\usepackage{cite}
\usepackage{graphicx}
\usepackage[cmex10]{amsmath}
\usepackage{algorithmic}
\usepackage{algorithm}
\usepackage{array}
\usepackage{mdwmath}
\usepackage{mdwtab}
\usepackage[caption=false]{caption}
\usepackage[caption=false,font=footnotesize]{subfig}
\usepackage{fixltx2e}
\usepackage{stfloats}
\ifCLASSOPTIONcaptionsoff
 \usepackage[nomarkers]{endfloat}
 \let\MYoriglatexcaption\caption
 \renewcommand{\caption}[2][\relax]{\MYoriglatexcaption[#2]{#2}}
\fi
\usepackage{url}

\begin{document}
%
\title{On the Cram\'{e}r-Rao Lower Bound for Frequency Correlation Matrices of Doubly Selective Fading Channels for OFDM Systems}

\author{\IEEEauthorblockN{Xiaochuan~Zhao, Ming~Yang, Tao~Peng and Wenbo~Wang}%
\IEEEauthorblockA{Wireless Signal Processing and Network Lab \\
Key Laboratory of Universal Wireless Communication, Ministry of Education \\
Beijing University of Posts and Telecommunications, Beijing, China \\
Email: zhaoxiaochuan@gmail.com}%
\thanks{This work is sponsored in
part by the National Natural Science Foundation of China under grant
No.60572120 and 60602058, and in part by the national high
technology researching and developing program of China (National 863
Program) under grant No.2006AA01Z257 and by the National Basic
Research Program of China (National 973 Program) under grant
No.2007CB310602.}}

\maketitle

\begin{abstract}
In this paper, the Cram\'er-Rao lower bound (CRLB) of the sample
frequency correlation matrices (SFCM) is derived based on a rigorous
model of the doubly selective fading channel for orthogonal
frequency division multiplexing (OFDM) systems with
pilot-symbol-aided modulation. By assuming a fixed pilot sequence
and independent samples, SFCM is complex Wishart distributed. Then,
the maximum likelihood estimator (MLE) and the exact expression of
CRLB are obtained. From CRLB, the lower bounds of total mean squared
error (TMSE) and average mean squared error (AvgMSE) independent of
the pilot sequence are deduced, which reveal that the amount of
samples is the dominant factor affecting AvgMSE while the
signal-to-noise ratio and the maximum Doppler spread have negligible
effect. Numerical simulations demonstrate the analytic results.
\end{abstract}

\begin{IEEEkeywords}
CRLB, Frequency correlation matrix, Doubly selective fading
channels, OFDM.
\end{IEEEkeywords}

\renewcommand{\thefootnote}{\fnsymbol{footnote}}

\section{Introduction}
\label{sec:intro}  Playing a key role in the channel estimation for
orthogonal frequency division multiplexing (OFDM) systems, the
frequency correlation matrix (FCM) is utilized by many
statistics-based channel estimation algorithms, e.g., the linear
minimum mean-squared error (LMMSE) estimator and its optimal
low-rank approximations \cite{Edfo98}, the MMSE estimator exploring
both time and frequency correlations \cite{Li98}, the
two-dimensional Wiener filtering \cite{Hoeh97}, and those algorithms
based on parametric channel model
\cite{Yang01}\cite{Ragh07}\cite{ZhaoPIMRC}. In real applications,
the sample FCM (SFCM) is used in stead of the true one, and usually
obtained through the least squared (LS) channel estimation.

For fixed or slowly moving radio channels whose Doppler spreads are
relatively small, the channels reveal a feature of block-fading
\cite{Shiu00}. Hence, the Doppler spread affects the accumulation of
SFCM negligibly. However, for fast moving radio channels,
intra-symbol fading becomes so prominent that inter-carrier
interference (ICI) takes effect by not only degrading the
performance of OFDM systems \cite{Wang06}, but also affecting the
mean and covariance of SFCM. In \cite{Li01}, the bounds of the ICI
power is derived.

The bias-property of SFCM in doubly selective fading channels has
been investigated in \cite{ZhaoICC09Bias}. As a counterpart, in this
paper, we find out the Cram\'er-Rao lower bound (CRLB) for FCM to
evaluate the performance of maximum likelihood estimator (MLE) and
uncover the factors influencing the estimation accuracy.

This paper is organized as follows. In Section \ref{sec:model}, the
OFDM system and channel model are introduced. Then, in Section
\ref{sec:CRLB}, CRLB for FCM is derived and further discussed to
uncover the essential factors. Numerical results appear in Section
\ref{sec:numresults}. Finally, Section \ref{sec:conclusion}
concludes the paper.

\emph{Notation}: Lowercase and uppercase boldface letters denote
column vectors and matrices, respectively. $(\cdot)^*$, $(\cdot)^H$,
and $||\cdot||_F$ denote conjugate, conjugate transposition, and
Frobenius norm, respectively. $\otimes$ denotes the Kronecker
product. $E(\cdot)$ represents expectation. $[{\bf{A}}]_{i,j}$ and
$[{\bf{a}}]_{i}$ denotes the $(i$,$j)$-th element of ${\bf{A}}$ and
the $i$-th element of ${\bf{a}}$, respectively.
${\text{diag}}({\bf{a}})$ is a diagonal matrix by placing ${\bf{a}}$
on the diagonal.

\setlength{\arraycolsep}{0.2em}

\section{System Model}
\label{sec:model} Consider an OFDM system with a bandwidth of
$BW=1/T$ Hz ($T$ is the sampling period). $N$ denotes the total
number of tones, and a cyclic prefix (CP) of length $L_{cp}$ is
inserted before each symbol to eliminate inter-block interference.
Thus the whole symbol duration is $T_s=(N+L_{cp})T$.

The complex baseband model of a linear time-variant mobile channel
with $L$ paths can be described by \cite{Stee92}
\begin{equation}
\label{eqn:channel}
{h(t,\tau)=\sum\limits_{l=1}^{L}{h_l(t)\delta\left({\tau-\tau_lT}\right)}}
\end{equation}
where $\tau_l\in{\mathcal{R}}$ is the normalized non-sample-spaced
delay of the $l$-th path, and $h_l(t)$ is the corresponding complex
amplitude. According to the wide-sense stationary uncorrelated
scattering (WSSUS) assumption, $h_l(t)$'s are modeled as
uncorrelated narrowband complex Gaussian processes.

Furthermore, by assuming the uniform scattering environment
introduced by Clarke \cite{Clar68}, $h_l(t)$'s have the identical
normalized time correlation function (TCF) for all $l$'s, thus the
TCF of the $l$'s path is
\begin{equation}
\label{eqn:rtdef}
{r_{t,l}({\Delta}t)=\sigma_l^2J_0\left(2{\pi}{f_d}{\Delta}t\right)}
\end{equation}
where $\sigma _l^2$ is the power of the $l$-th path, $f_d$ is the
maximum Doppler spread, and $J_0(\cdot)$ is the zeroth order Bessel
function of the first kind. Additionally we assume the power of
channel is normalized, i.e.,
$\sum\nolimits_{l=0}^{L-1}\sigma_l^2=1$.

Assuming a sufficient CP, i.e., $L_{cp} \geq L$, the discrete signal
model in the frequency domain is written as
\begin{equation}
\label{eqn:YmMatrixdef}
{\bf{y}}_f(n)={\bf{H}}_f(n){\bf{x}}_f(n)+{\bf{n}}_f(n)
\end{equation}
where
${\bf{x}}_f(n),{\bf{y}}_f(n),{\bf{n}}_f(n)\in{\mathcal{C}}^{N\times1}$
are the $n$-th transmitted and received signal and additive white
Gaussian noise (AWGN) vectors, respectively, and
${\bf{H}}_f(m)\in{\mathcal{C}}^{N\times{N}}$ is the channel transfer
matrix with the $(k+\nu,k)$-th element as
\begin{equation}
\label{eqn:Hdef}
\left[{\bf{H}}_f(n)\right]_{k+\upsilon,k}=\frac{1}{N}\sum\limits_{m=0}^{N-1}\sum\limits_{l=1}^{L}h_l(n,m)e^{-j2{\pi}({\upsilon}m+k{\tau_l})/N}
\end{equation}
where $h_l(n,m)=h_l(nT_s+(L_{cp}+m)T)$ is the sampled complex
amplitude of the $l$-th path. $k$ and $\upsilon$ denote frequency
and Doppler spread, respectively. Apparently, as ${\bf{H}}_f(n)$ is
non-diagonal, ICI is present. In fact, when the normalized maximum
Doppler spread $f_dT_s\le0.1$, the signal-to-interference ratio
(SIR) is over 17.8 dB \cite{Choi01}.

\section{CRLB for Frequency Correlation Matrices}
\label{sec:CRLB}  Usually SFCM is obtained through the LS channel
estimation. We consider OFDM systems adopting pilot-symbol-assisted
modulation (PSAM) \cite{Edfo98}, hence only pilot symbols, denoted
as ${\bf{y}}_{p}(n)\in\mathcal{C}^{N\times1}$, are extracted and
used to perform LS channel estimation. In addition, the pilot
sequence is assumed to be invariant along the time. Therefore,
\begin{equation}
\label{eqn:lsdef}
{\bf{h}}_{p,ls}(n)={\bf{X}}_{p}^{-1}{\bf{y}}_{p}(n)={\bf{X}}_{p}^{-1}{\bf{H}}_{p}(n){\bf{x}}_{p}+{\bf{X}}_{p}^{-1}{\bf{n}}_{p}(n)
\end{equation}
where ${\bf{X}}_{p}={\text{diag}}({\bf{x}}_{p})$ is a diagonal
matrix consisting of pilot symbols, and the noise term is
${\bf{n}}_p(n)\sim{\mathcal{CN}}(0,\sigma_n^2{\bf{I}}_N)$.

Denote the instantaneous channel impulse response (CIR) vector as
${\bf{h}}_t(n,m)=[h_1(n,m),\ldots,h_L(n,m)]^T$, $m=0,\ldots,N-1$,
according to the assumptions of WSSUS and uniform scattering,
${\bf{h}}_t(n,m)$ is complex normal, i.e.,
\begin{equation}
{\bf{h}}_t(n,m)\;{\sim}\;{\mathcal{CN}}_L(0,{\bf{D}})\nonumber
\end{equation}
where ${\bf{D}}={\text{diag}}({\sigma}_l^2)$, $l=1,\ldots,L$. Then
form the CIR matrix as
${\bf{H}}_t(n)=[{\bf{h}}_t(n,0),\ldots,{\bf{h}}_t(n,N-1)]$, so we
have
\begin{equation}
{\text{vec}}\left({\bf{H}}_t(n)\right)\;{\sim}\;{\mathcal{CN}}_{LN}(0,{\bf{\Omega}}\otimes{\bf{D}})\nonumber
\end{equation}
where ${\bf{\Omega}}\in{\mathcal{C}}^{N\times{N}}$ is a Toeplitz
time correlation matrix (TCM), defined as
\begin{equation}
\label{eqn:Gammedef}
[{\bf{\Omega}}]_{m_1,m_2}=J_0\left(2{\pi}{f_d}(m_1-m_2)T\right)
\end{equation}
Then according to (\ref{eqn:Hdef}), the channel transfer matrix
${\bf{H}}_f(n)={\bf{F}}_{\tau}{\bf{H}}_t(n)$, where
${\bf{F}}_{\tau}\in{\mathcal{C}}^{N\times{L}}$ is the unbalanced
Fourier transform matrix, defined as
$[{\bf{F}}_{\tau}]_{k,l}=e^{-j2{\pi}k{\tau}_l/N}$. Thus
\begin{equation}
\label{pdf:Hf}
{\bf{H}}_f(n)\;{\sim}\;{\mathcal{CN}}_{N\times{N}}(0,{\bf{\Omega}}\otimes({\bf{F}}_{\tau}{\bf{D}}{\bf{F}}_{\tau}^H))
\end{equation}
Assuming CIR is independent of the thermal noise, with
(\ref{eqn:lsdef}) and (\ref{pdf:Hf}), we have
\begin{equation}
\label{pdf:hls}
{\bf{h}}_{p,ls}(n)\;{\sim}\;{\mathcal{CN}}_{N}(0,{\bf{\Sigma}})
\end{equation}
where the covariance matrix ${\bf{\Sigma}}$ is defined as
\begin{equation}
\label{eqn:Sigmadef}
{\bf{\Sigma}}=\omega{\bf{X}}_{p}^{-1}({\bf{R}}_{p}+\frac{{\sigma}_n^2}{\omega}{\bf{I}}_{N}){\bf{X}}_{p}^{-H}
\end{equation}
where $\omega={\bf{x}}_{p}^H{\bf{\Omega}}{\bf{x}}_{p}$, and
${\bf{R}}_{p}={\bf{F}}_{\tau}{\bf{D}}{\bf{F}}_{\tau}^H$ is the true
FCM.

When the LS estimated CFR's, i.e., ${\bf{h}}_{p,ls}(n)$'s, are
available, SFCM is constructed as
\begin{equation}
\label{eqn:estRdef}
{\bf{\hat{R}}}_{p,ls}=\frac{1}{N_{t}}\sum\limits_{n=1}^{N_{t}}{\bf{h}}_{p,ls}(n){\bf{h}}_{p,ls}^H(n)
\end{equation}
where $N_t$ is the amount of samples. To derive the probability
density function (PDF) of SFCM, we assume that samples are
independent of each other, which may be a strict constraint.
However, when the maximum Doppler spread is large and the spacing
between two contiguous pilot symbols is comparatively small, the
correlation between them is rather low, which alleviates the effect
of model mismatch. Then, based on the assumption of independence and
(\ref{pdf:hls}), we know that SFCM has the complex central Wishart
distribution with $N_t$ degrees of freedom and covariance matrix
${\bf{\Sigma}}'={\bf{\Sigma}}/N_t$ \cite{Ratn03}, denoted as
\begin{equation}
\label{pdf:Rls}
{\bf{\hat{R}}}_{p,ls}\;{\sim}\;{\mathcal{CW}}_{N}(N_t,{\bf{\Sigma}}')
\end{equation}
and its PDF is
\begin{equation}
\label{pdf:Rlsfunc}
f({\bf{\hat{R}}}_{p,ls})=\frac{{\text{etr}}(-{\bf{\Sigma}}'^{-1}{\bf{\hat{R}}}_{p,ls})(\det({\bf{\hat{R}}}_{p,ls}))^{N_t-N}}{{C\Gamma}_N(N_t)(\det({\bf{\Sigma}}')^{N_t}}
\end{equation}
where ${\text{etr}}(\cdot)=\exp({\text{tr}}(\cdot))$ and
${C\Gamma}_N(N_t)$ is the complex multivariate gamma function,
defined as
\begin{equation}
\label{eqn:complexGamma}
{C\Gamma}_N(N_t)={\pi}^{N(N-1)/2}\prod\limits_{k=1}^{N}{\Gamma}(N_t-k+1)\nonumber
\end{equation}
Then, from (\ref{pdf:Rlsfunc}), the likelihood function is written
as
\begin{eqnarray}
{\text{L}}({\bf{R}}_{p})&=&{\text{tr}}(-{\bf{\Sigma}}'^{-1}{\bf{\hat{R}}}_{p,ls})+(N_t-N)\ln(\det({\bf{\hat{R}}}_{p,ls}))\nonumber\\
\label{likelihood:Rlsfunc}
{}&{}&-\ln({C\Gamma}_N(N_t))-N_t\ln(\det({\bf{\Sigma}}'))\nonumber
\end{eqnarray}
Therefore, the score function with respect to the parameter matrix
${\bf{R}}_{p}$ is
\begin{equation}
\label{score:Rlsfunc}
{\text{score}}({\bf{R}}_{p})=\frac{\partial{{\text{L}}({\bf{R}}_{p})}}{\partial{{\text{vec}}({\bf{R}}_{p})}}=\frac{\partial{{\text{vec}}({\bf{\Sigma}}')^T}}{\partial{{\text{vec}}({\bf{R}}_{p})}}\times\frac{\partial{{\text{L}}({\bf{R}}_{p})}}{\partial{{\text{vec}}({\bf{\Sigma}}')}}
\end{equation}
where the first term on the right-hand side of (\ref{score:Rlsfunc})
is
\begin{equation}
\label{eqn:score1stterm}
\frac{\partial{{\text{vec}}({\bf{\Sigma}}')^T}}{\partial{{\text{vec}}({\bf{R}}_{p})}}=\frac{\omega}{N_t}({\bf{X}}_{p}^{-H}\otimes{\bf{X}}_{p}^{-1})
\end{equation}
and the second term is
\begin{equation}
\label{eqn:score2ndterm}
\frac{\partial{{\text{L}}({\bf{R}}_{p})}}{\partial{{\text{vec}}({\bf{\Sigma}}')}}={\text{vec}}[({\bf{\Sigma}}'^{-1}{\bf{\hat{R}}}_{p,ls}{\bf{\Sigma}}'^{-1}-N_t{\bf{\Sigma}}'^{-1})^T]
\end{equation}

By letting the score function equal zero and with
(\ref{eqn:Sigmadef}), the MLE of FCM is derived as
\begin{equation}
\label{eqn:MLfirstdef}
{\text{MLE}}({\bf{R}}_{p})=\frac{{\bf{X}}_{p}{\bf{\hat{R}}}_{p,ls}{\bf{X}}_{p}^H-{\sigma}_n^2{\bf{I}}_N}{{\bf{x}}_{p}^H{\bf{\Omega}}{\bf{x}}_{p}}
\end{equation}
Note that (\ref{eqn:MLfirstdef}) relies on the pre-known TCM, i.e.,
${\bf{\Omega}}$, and noise power. For Rayleigh fading channels, it
means the maximum Doppler spread, $f_d$, is known.

\begin{figure}[!t]
\centering
\includegraphics[width=3.4in,height=2.4in]{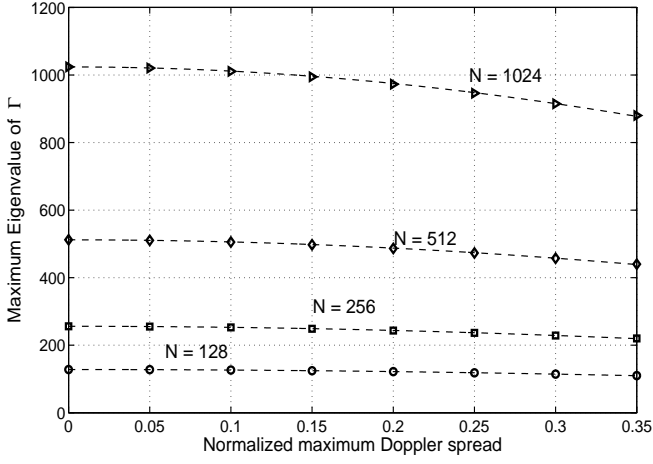}
\caption{Fitting the maximum eigenvalues of ${\bf{\Omega}}$ with
(\ref{eqn:fittingfunction}) for different sizes of FFT ($N$) and
normalized Doppler's ($f_dT_s$).} \label{fig:fig1}
\end{figure}

Further, according to the score function, the Fisher Information
matrix with respect to ${\bf{R}}_{p}$ is
\begin{equation}
\label{eqn:fisherinfordef}
{\bf{J}}({\bf{R}}_{p})=E\left[\left(\frac{\partial{{\text{L}}({\bf{R}}_{p})}}{\partial{{\text{vec}}({\bf{R}}_{p})}}\right)\left(\frac{\partial{{\text{L}}({\bf{R}}_{p})}}{\partial{{\text{vec}}({\bf{R}}_{p})}}\right)^H\right]
\end{equation}
With
(\ref{score:Rlsfunc})(\ref{eqn:score1stterm})(\ref{eqn:score2ndterm}),
we have
\begin{eqnarray}
\frac{\partial{{\text{L}}({\bf{R}}_{p})}}{\partial{{\text{vec}}({\bf{R}}_{p})}}&=&\frac{\omega}{N_t}({\bf{X}}_{p}^{-H}\otimes{\bf{X}}_{p}^{-1})\nonumber\\
{}&{}&\times{\text{vec}}[({\bf{\Sigma}}'^{-1}{\bf{\hat{R}}}_{p,ls}{\bf{\Sigma}}'^{-1}-N_t{\bf{\Sigma}}'^{-1})^T]\nonumber
\end{eqnarray}
so, (\ref{eqn:fisherinfordef}) is rewritten into
(\ref{eqn:fisherinfordefnew}), shown at the bottom of the next page.
\begin{figure*}[!b]
\normalsize \vspace{2pt} \hrule \normalsize {\begin{equation}
\label{eqn:fisherinfordefnew}
{\bf{J}}({\bf{R}}_{p})=\frac{\omega^2}{N_t^2}({\bf{X}}_{p}^{-H}\otimes{\bf{X}}_{p}^{-1})E\{{\text{vec}}[({\bf{\Sigma}}'^{-1}{\bf{\hat{R}}}_{p,ls}{\bf{\Sigma}}'^{-1}-N_t{\bf{\Sigma}}'^{-1})^T]{\text{vec}}[({\bf{\Sigma}}'^{-1}{\bf{\hat{R}}}_{p,ls}{\bf{\Sigma}}'^{-1}-N_t{\bf{\Sigma}}'^{-1})^T]^H\}({\bf{X}}_{p}^{-H}\otimes{\bf{X}}_{p}^{-1})^H
\end{equation}}
\end{figure*}
Notice that
\begin{equation}
{\bf{\Sigma}}'^{-1}{\bf{\hat{R}}}_{p,ls}{\bf{\Sigma}}'^{-1}\;{\sim}\;{\mathcal{CW}}_{N}(N_t,{\bf{\Sigma}}'^{-1})\nonumber
\end{equation}
and
\begin{equation}
E[{\bf{\Sigma}}'^{-1}{\bf{\hat{R}}}_{p,ls}{\bf{\Sigma}}'^{-1}]=N_t{\bf{\Sigma}}'^{-1}\nonumber
\end{equation}
therefore
\begin{eqnarray}
{}&{}&E\{{\text{vec}}[({\bf{\Sigma}}'^{-1}{\bf{\hat{R}}}_{p,ls}{\bf{\Sigma}}'^{-1}-N_t{\bf{\Sigma}}'^{-1})^T]\nonumber\\
{}&{}&\;\;\;\;\;\;\;\;\;\;\;\;\times{\text{vec}}[({\bf{\Sigma}}'^{-1}{\bf{\hat{R}}}_{p,ls}{\bf{\Sigma}}'^{-1}-N_t{\bf{\Sigma}}'^{-1})^T]^H\}\nonumber\\
\label{eqn:secondterm}
{}&=&{\text{Var}}\{{\text{vec}}[({\bf{\Sigma}}'^{-1}{\bf{\hat{R}}}_{p,ls}{\bf{\Sigma}}'^{-1})^T]\}
\end{eqnarray}
Given ${\bf{S}}\sim{\mathcal{CW}}_{N}(N_t,{\bf{\Sigma}}')$, the
entry of its second origin moment is \cite{Maiw00}
\begin{equation}
E([{\bf{S}}]_{i,j}[{\bf{S}}]_{k,l})=N_t^2[{\bf{\Sigma}}']_{i,j}[{\bf{\Sigma}}']_{k,l}+N_t[{\bf{\Sigma}}']_{k,j}[{\bf{\Sigma}}']_{i,l}\nonumber
\end{equation}
Therefore, the entry of its second central moment is
\begin{equation}
E[([{\bf{S}}]_{i,j}-E([{\bf{S}}]_{i,j}))([{\bf{S}}]_{k,l}-E([{\bf{S}}]_{k,l}))]=N_t[{\bf{\Sigma}}']_{k,j}[{\bf{\Sigma}}']_{i,l}\nonumber
\end{equation}
Accordingly, (\ref{eqn:secondterm}) is rewritten as
\begin{equation}
\label{eqn:vardef}
{\text{Var}}\{{\text{vec}}[({\bf{\Sigma}}'^{-1}{\bf{\hat{R}}}_{p,ls}{\bf{\Sigma}}'^{-1})^T]\}=N_t({\bf{\Sigma}}'^{-H}\otimes{\bf{\Sigma}}'^{-T})
\end{equation}
Then, with (\ref{eqn:vardef}), ${\bf{J}}({\bf{R}}_{p})$ is
\begin{equation}
{\bf{J}}({\bf{R}}_{p})=\frac{\omega^2}{N_t}({\bf{X}}_{p}^{-H}\otimes{\bf{X}}_{p}^{-1})({\bf{\Sigma}}'^{-H}\otimes{\bf{\Sigma}}'^{-T})({\bf{X}}_{p}^{-1}\otimes{\bf{X}}_{p}^{-H})\nonumber
\end{equation}

From the Fisher Information matrix, the CRLB of ${\bf{R}}_{p}$ can
be derived as \cite{Mard79}\cite{Golu96}
\begin{eqnarray}
{\text{CRLB}}({\bf{R}}_{p})&=&{\bf{J}}^{-1}({\bf{R}}_{p})\nonumber\\
{}&=&\frac{N_t}{\omega^2}({\bf{X}}_{p}\otimes{\bf{X}}_{p}^{H})({\bf{\Sigma}}'^H\otimes{\bf{\Sigma}}'^T)({\bf{X}}_{p}^{H}\otimes{\bf{X}}_{p})\nonumber\\
\label{eqn:CRLBdef}
{}&=&\frac{1}{N_t}(\frac{1}{\omega}{\bf{X}}_{p}{\bf{\Sigma}}^H{\bf{X}}_{p}^{H})\otimes(\frac{1}{\omega}{\bf{X}}_{p}^{H}{\bf{\Sigma}}^T{\bf{X}}_{p})
\end{eqnarray}
With (\ref{eqn:Sigmadef}), (\ref{eqn:CRLBdef}) can be further
written as
\begin{equation}
\label{eqn:CRLBdefMPSKnew}
{\text{CRLB}}({\bf{R}}_{p})=\frac{1}{N_t}({\bf{R}}_{p}+\frac{{\sigma}_n^2}{\omega}{\bf{I}}_{N})\otimes({\bf{R}}_{p}+\frac{{\sigma}_n^2}{\omega}{\bf{I}}_{N})^T
\end{equation}

Based on (\ref{eqn:CRLBdefMPSKnew}), a lower bound of the total mean
squared error (TMSE) for ${\text{MLE}}({\bf{R}}_{p})$ is
\begin{eqnarray}
{\text{TMSE}}_{LB}({\bf{R}}_{p})&=&{\text{tr}}({\text{CRLB}}({\bf{R}}_{p}))\nonumber\\
{}&=&\frac{1}{N_t}{\text{tr}}^2({\bf{R}}_{p}+\frac{{\sigma}_n^2}{\omega}{\bf{I}}_{N})\nonumber\\
\label{eqn:TMSELBdef}
{}&=&\frac{N^2}{N_t}(1+\frac{1}{\omega\gamma})^2
\end{eqnarray}
where $\gamma=\sigma_n^{-2}$ is the signal-to-noise ratio (SNR).
And, accordingly, the lower bound of the average mean squared error
(avgMSE) is
\begin{equation}
\label{eqn:AvgMSELBdef}
{\text{AvgMSE}}_{LB}({\bf{R}}_{p})=\frac{{\text{TMSE}}_{LB}({\bf{R}}_{p})}{N^2}=\frac{1}{N_t}(1+\frac{1}{\omega\gamma})^2
\end{equation}
(\ref{eqn:AvgMSELBdef}) verifies the common sense that the more
samples collected, the more accurate estimation acquired. And it
also reveals that increasing SNR can reduce the estimation error.
Furthermore, since
\begin{equation}
\label{eqn:omegadefnew}
{\omega}={\bf{x}}_p^H{\bf{\Omega}}{\bf{x}}_p=\|{\bf{x}}_p\|_2^2\times\frac{{\bf{x}}_p^H{\bf{\Omega}}{\bf{x}}_p}{{\bf{x}}_p^H{\bf{x}}_p}=\|{\bf{x}}_p\|_2^2\times{\text{R}}_{{\bf{x}}_{p}}({\bf{\Omega}})
\end{equation}
where ${\text{R}}_{{\bf{x}}_{p}}({\bf{\Omega}})$ is the Rayleigh
quotient of ${\bf{\Omega}}$ associated with the pilot sequence
${\bf{x}}_p$, and
${\text{R}}_{{\bf{x}}_{p}}({\bf{\Omega}})\leq{\lambda}_{max}$ where
${\lambda}_{max}$ is the maximum eigenvalue of ${\bf{\Omega}}$.
Besides, when the power of pilot symbol is normalized,
$\|{\bf{x}}_p\|_2^2=N$. Hence (\ref{eqn:AvgMSELBdef}) is further
lower bounded by
\begin{equation}
\label{eqn:AvgMSELBdefnew}
{\overline{\text{AvgMSE}}}_{LB}({\bf{R}}_{p})=\frac{1}{N_t}(1+\frac{1}{N{\lambda}_{max}\gamma})^2
\end{equation}
To further look into the relationship between $f_dT_s$ and
${\lambda}_{max}$, we examine the extreme eigenvalues of
${\bf{\Omega}}$ for different $f_dT_s$'s and $N$'s numerically, and
the results are plotted in Fig.\ref{fig:fig1}. Moreover, we find a
simple function fitting the maximum eigenvalues of all cases very
well. The function is
\begin{equation}
\label{eqn:fittingfunction}
{\lambda}_{max}({\bf{\Omega}})=NJ_0(2{\pi}cf_dT_s)
\end{equation}
where $c=0.35$ when $f_dT_s\le0.35$\footnote{This condition ensures
that $J_0(2{\pi}{\alpha}f_dT_s)$ is positive and monotonically
decreasing with respect to $f_dT_s$. In fact, this condition is
always satisfied since current applied OFDM systems have
$f_dT_s\le0.1$ to maintain the power of ICI within a tolerable range
\cite{Choi01}.}. Therefore, a more insightful lower bound can be
achieved by
\begin{equation}
\label{eqn:AvgMSELBdefnewapprox}
{\overline{\text{AvgMSE}}}_{LB}({\bf{R}}_{p})=\frac{1}{N_t}(1+\frac{1}{N^2J_0(2{\pi}{c}f_dT_s)\gamma})^2
\end{equation}
According to (\ref{eqn:AvgMSELBdefnewapprox}), we know that the
amount of samples, i.e., $N_t$, effects the estimation accuracy
dominantly but SNR and maximum Doppler spread do not, since $N^2$ is
sufficiently large for most of current systems.

\section{Numerical Results}
\label{sec:numresults} The OFDM system in simulations is of
$BW=1.25$ MHz ($T=1/BW=0.8$ ms), $N=128$, and $L_{cp}=16$. Two 3GPP
E-UTRA channel models are adopted: Extended Vehicular A model (EVA)
and Extended Typical Urban model (ETU) \cite{3GPP36101}. The excess
tap delay of EVA is [$0$, $30$, $150$, $310$, $370$, $710$, $1090$,
$1730$, $2510$] ns, and its relative power is [$0.0$, $-1.5$,
$-1.4$, $-3.6$, $-0.6$, $-9.1$, $-7.0$, $-12.0$, $-16.9$] dB. For
ETU, they are [$0$, $50$, $120$, $200$, $230$, $500$, $1600$,
$2300$, $5000$] ns and [$-1.0$, $-1.0$, $-1.0$, $0.0$, $0.0$, $0.0$,
$-3.0$, $-5.0$, $-7.0$] dB, respectively. The classic Doppler
spectrum, i.e., Jakes' spectrum \cite{Stee92}, is applied to
generate the Rayleigh fading channel.

\begin{figure}[!t]
\centering
\includegraphics[width=3.4in,height=2.4in]{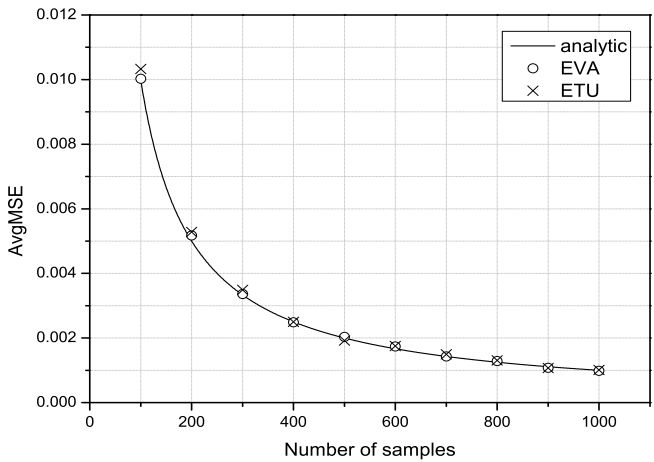}
\caption{Comparison of analytic results (\ref{eqn:AvgMSELBdef}) and
numerical results for EVA and ETU channels when $\gamma=20$dB and
$f_d=200$Hz.} \label{fig:fig2}
\end{figure}

In Fig.\ref{fig:fig2}, we compare the analytic results
(\ref{eqn:AvgMSELBdef}) and the numerical results over a range of
$N_t$'s for EVA and ETU channels, respectively, when $\gamma=20$dB
and $f_d=200$Hz. The pilot sequences are QPSK modulated and randomly
chosen. And the collected samples are apart from each others far
enough to guarantee the assumption of independence. Apparently, the
analytic results meet the numerical ones quite well.

\begin{figure}[!t]
\centering
\includegraphics[width=3.4in,height=2.4in]{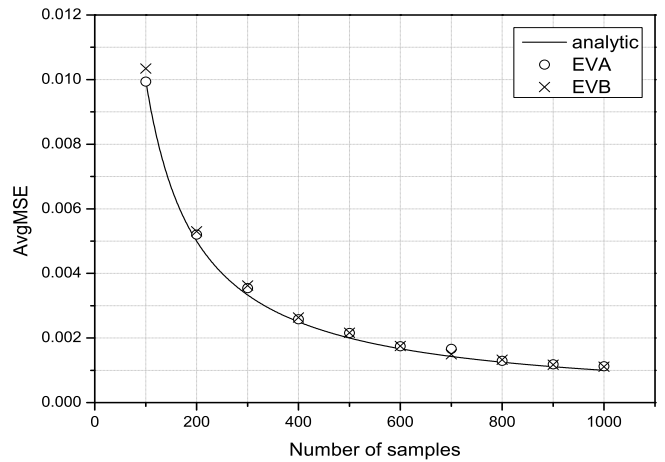}
\caption{Comparison of analytic results
(\ref{eqn:AvgMSELBdefnewapprox}) and numerical results for EVA and
ETU channels when $\gamma=20$dB and $f_d=200$Hz.} \label{fig:fig3}
\end{figure}

In Fig.\ref{fig:fig3}, we compare the analytic results
(\ref{eqn:AvgMSELBdefnewapprox}) and the numerical results for EVA
and ETU channels, respectively, when $\gamma=20$dB and $f_d=200$Hz.
The pilot sequences are QPSK modulated. In order to examine the
effect of different pilot sequences on $\omega$, one hundred
different sequences randomly generated are tested and their MSE's
are averaged and plotted. From the figure, we find that
(\ref{eqn:AvgMSELBdefnewapprox}) is a tight bound even for an
arbitrary pilot sequence.

\begin{figure*}[!t]
\centering{\subfloat[EVA,$f_d=200$Hz.]{\includegraphics[width=3.4in,height=2.4in]{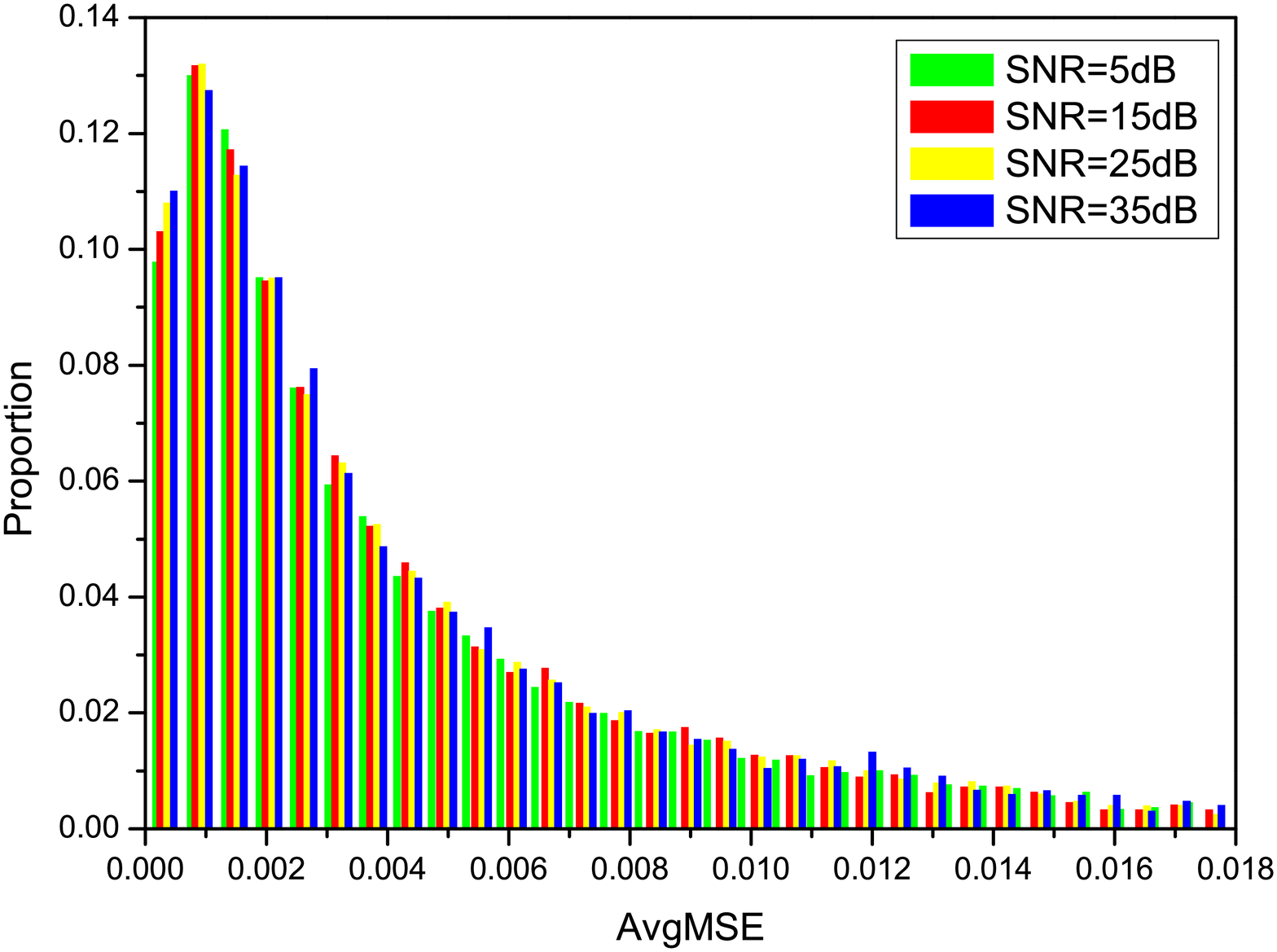}
\label{subfig:evasnr}} \hfil
\subfloat[ETU,$f_d=200$Hz.]{\includegraphics[width=3.4in,height=2.4in]{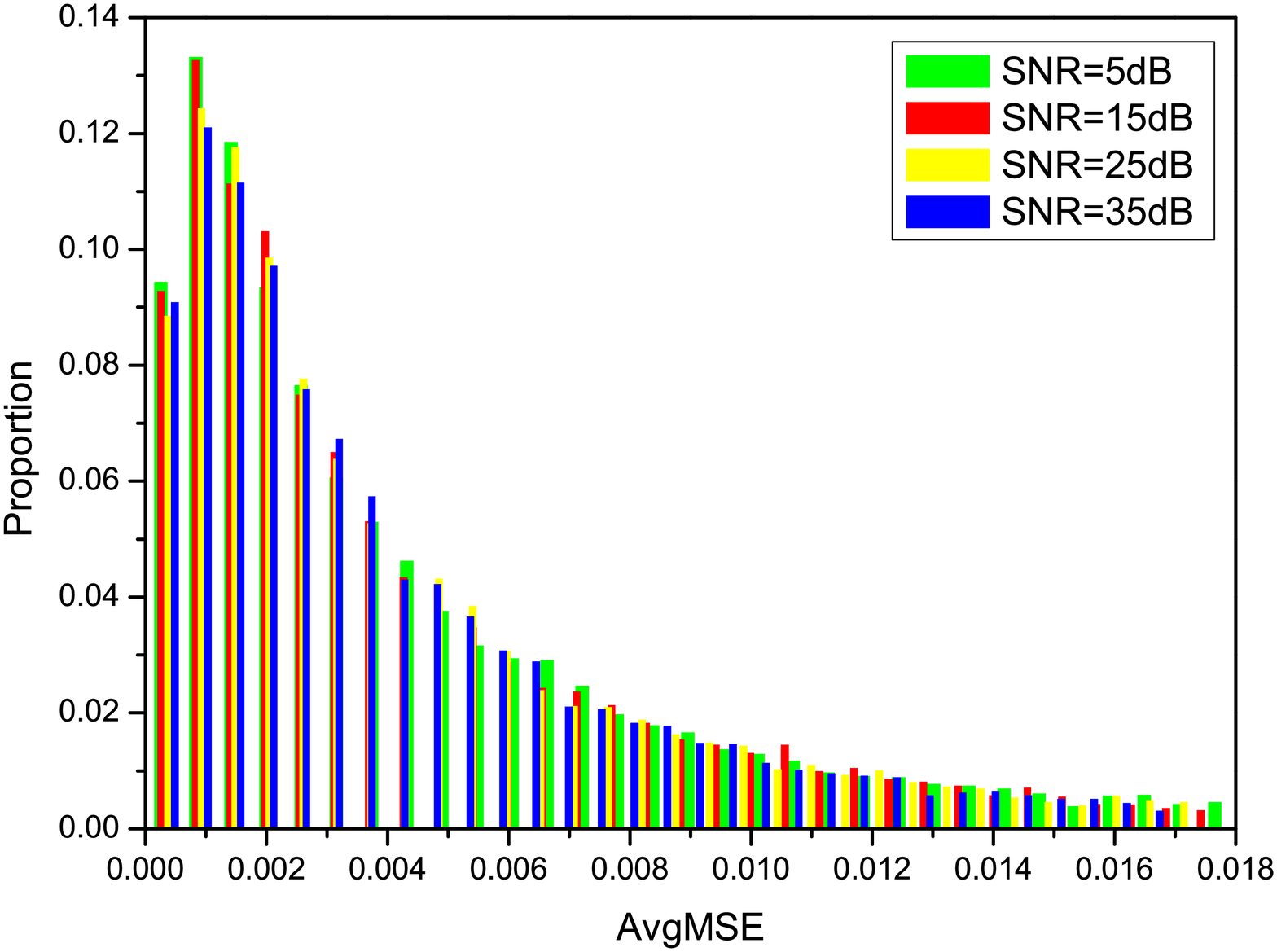}
\label{subfig:etusnr}}} \\
\centering{\subfloat[EVA,$\gamma=15$dB.]{\includegraphics[width=3.4in,height=2.4in]{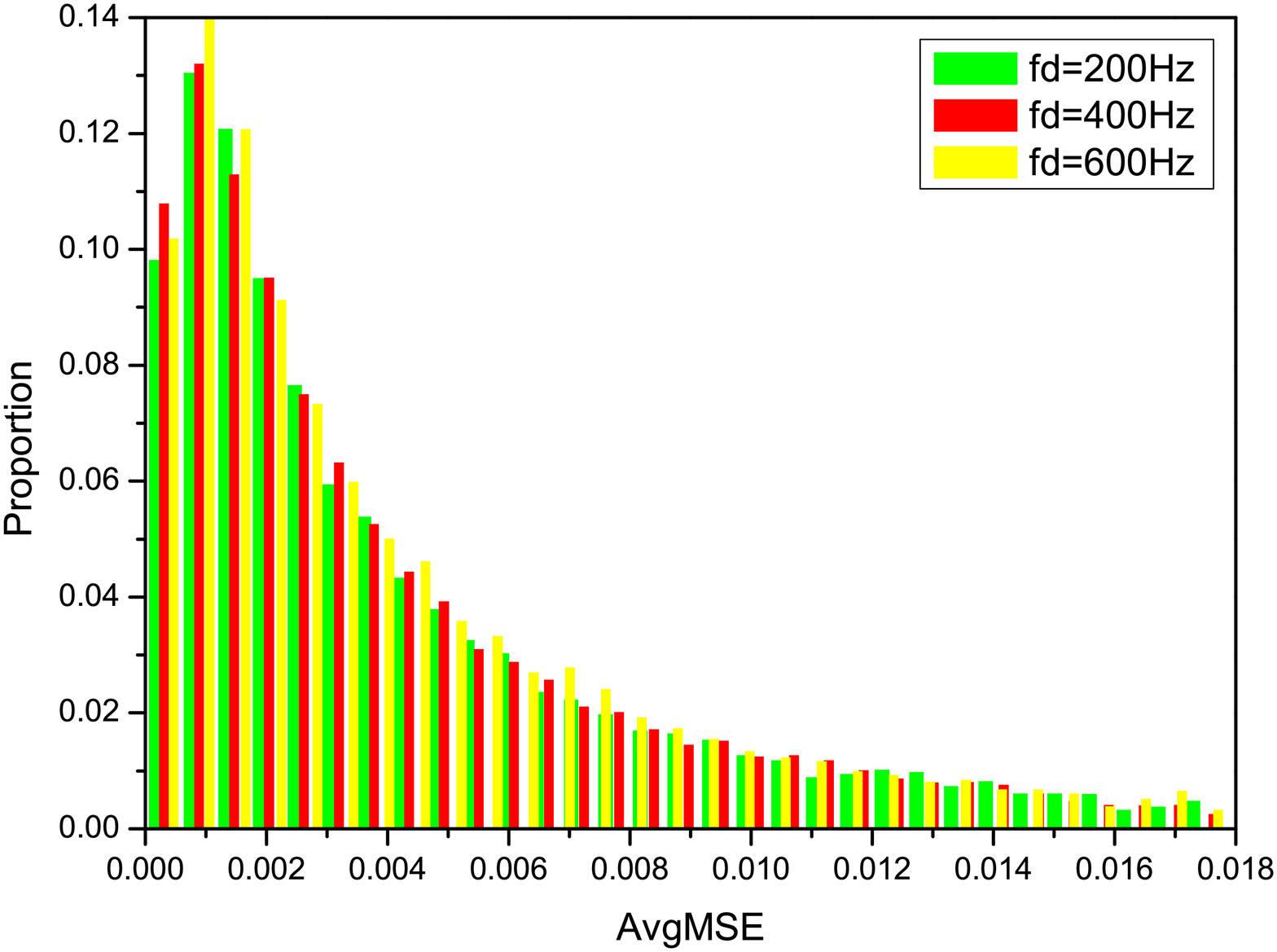}
\label{subfig:evadop}} \hfil
\subfloat[ETU,$\gamma=15$dB.]{\includegraphics[width=3.4in,height=2.4in]{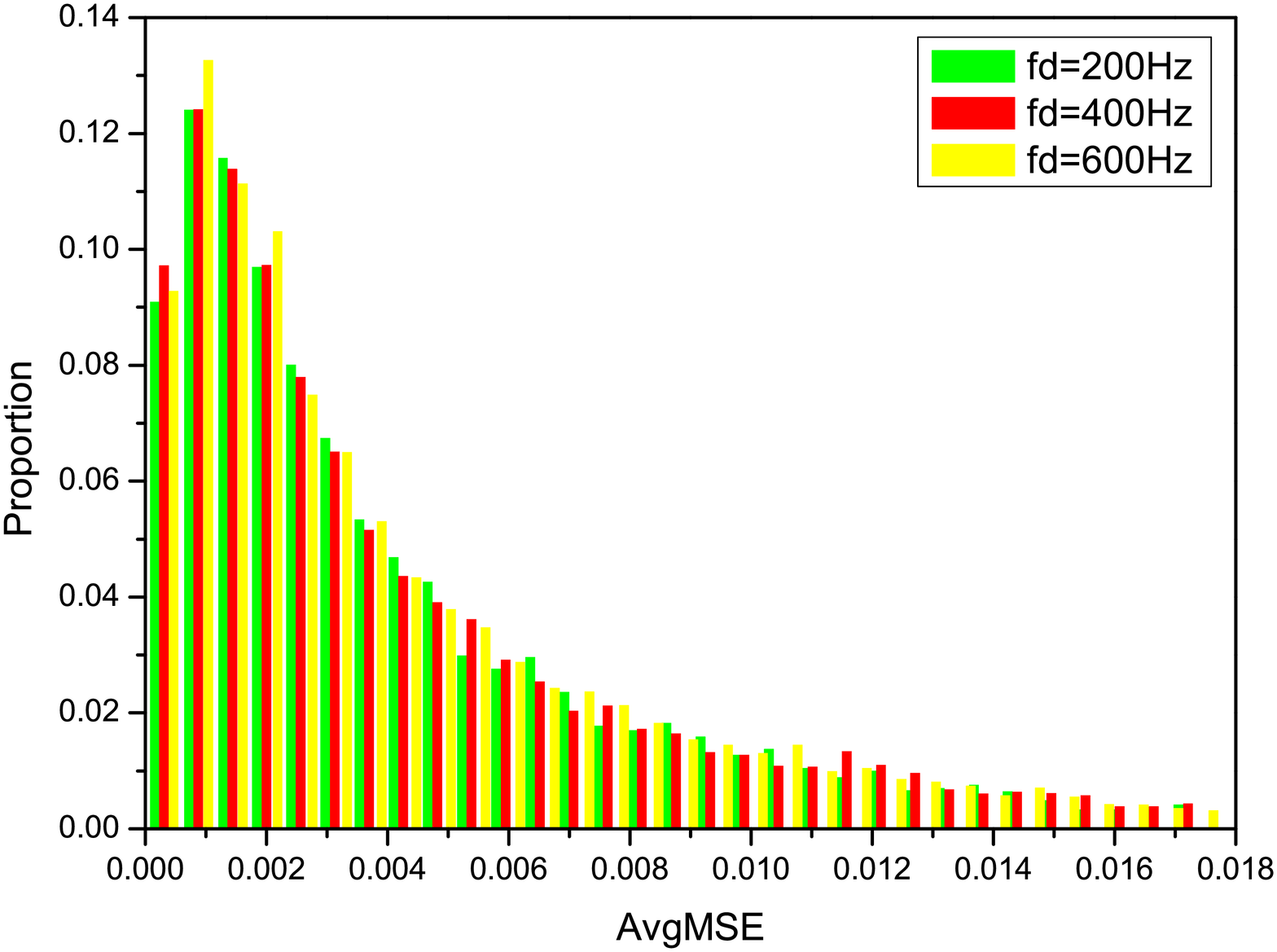}
\label{fig_fourth_case}}} \caption{Distributions of average MSE for
EVA and ETU channels when $N_t=200$.} \label{fig:fig4}
\end{figure*}

The distributions of avgMSE for different SNR's and Doppler's are
plotted in Fig.\ref{fig:fig4} through ten thousands estimations for
EVA and ETU channels, respectively. The amount of samples of each
test is 200, and the pilot sequences are QPSK modulated and randomly
generated. Clearly, avgMSE's are centered around zero and most of
them are within the range of zero to CRLB, which follows that
(\ref{eqn:MLfirstdef}) is an unbiased estimator. Moreover, it is
also obvious that the distributions of avgMSE for EVA and ETU
channels are negligibly influenced by $\gamma$ and $f_d$, which
follows the analytic lower bound (\ref{eqn:AvgMSELBdefnewapprox}).

\section{Conclusion}
\label{sec:conclusion} In this paper, the maximum likelihood
estimator and CRLB of the frequency correlation matrix for OFDM
systems in doubly selective fading channels are derived and
analyzed. Through the analyses, we obtain an insightful lower bound
of average MSE, i.e., (\ref{eqn:AvgMSELBdefnewapprox}), and
according to which, the amount of samples shows a dominant impact on
the accuracy of estimation while SNR and maximum Doppler spread have
relatively small effect when the number of subcarriers are
sufficiently large, although increasing SNR and decreasing maximum
Doppler spread can help to reduce MSE slightly.

\setlength{\arraycolsep}{5pt}

\bibliographystyle{IEEEtran}
\bibliography{IEEEabrv,myBibs}

\end{document}